\documentclass{elsart5p}

\usepackage{amsmath,amssymb,bbm,graphicx}

\begin{document}
\begin{frontmatter}

\title{On the Security of the Ping-Pong Protocol}

\author[kim]{Kim Bostr\"om},
\author[timo]{Timo Felbinger}

\address[kim]{Psychologisches Institut II, Universit\"at M\"unster, 48149 M\"unster, Germany}
\address[timo]{Institut f\"ur Physik, Universit\"at Potsdam, 14469 Potsdam, Germany}

\begin{abstract}
We briefly review the security of the ping-pong protocol in light of several
attack scenarios suggested by various authors since the proposal of the
protocol. We refute one recent attack on an ideal quantum channel, and
show that a recent claim of falseness of our original security proof is
erroneous.
\end{abstract}


\end{frontmatter}

\section{Introduction}

It is now five years that we proposed a quantum cryptographic protocol~\cite{BostroemFelbinger2002} whose novel feature was that the bits are transmitted in a deterministic manner: Alice, the sender, determines the bit value decoded by Bob, the receiver.  Thereby, the transmission efficiency is doubled as compared to non-deterministic protocols like the BB84~\cite{BennettBrassard1984}, where only 50\% of the transmitted bits can be used for communication purposes.  We were able to rigorously prove that in the case of a perfect quantum channel any effective eavesdropping attack can be detected.  Specifically, when Eve, the eavesdropper, tries to gain full information, that is $I_0=1$ bit per message bit, then the detection probability reads $d=1/2$ per control bit.  For comparison, in a similar scenario with Eve fully attacking all transmitted bits, the BB84 protocol provides a detection probability of $d=1/4$ per control bit.

Due to its deterministic nature, the protocol allows two parties to communicate
directly in a secure manner.  More precisely, the direct communication is
\emph{quasi-secure}, which means that the probability for an eavesdropper to
remain undetected declines exponentially with the length of the message
transmitted.  Alternatively, when perfect asymptotic security is required, the
protocol can be used as a quantum key distribution scheme: The sender
transmits a meaningless stream of random bits upon which the usual techniques
of error correction and privacy amplification can be applied so
that a shared secret key is established for later use in classical encryption. 

One peculiarity of the protocol is that the carrier of information, a single
qubit, is travelling forth and back between sender and receiver, which
motivated the name \emph{ping-pong protocol}. Another peculiarity is that
sender and receiver randomly switch between two modes: \emph{message mode} and
\emph{control mode}. Only in message mode a message bit is transferred and
only in control mode an eavesdropper can be detected with a certain probability.
The fact that the mode
can only be noticed by the eavesdropper when it is too late to escape
detection, is an important ingredient to the security of the protocol and
requires careful attention in experimental implementations.

One experimental implementation of the ping-pong protocol using entangled photons has been accomplished at the quantum optics lab in Potsdam, Germany, where the random switching between message mode and control mode is realized in an elegant way~\cite{OstermeyerWalenta2007}. 

Also meanwhile, several alternative quantum
cryptographic protocols have been proposed that are tailored along the scheme
of the original ping-pong protocol, providing improved efficiency or experimental
feasibility~\cite{CaiLi2004,CaiLi2004a,DegiovanniRuo-Berchera2004,LucamariniMancini2005,DengLong2004}. Since the security of these protocols is
based on the security of the original one, we would like to give a brief review on
the security situation of the ping-pong protocol as it appears nowadays, after
several attempts to attack the protocol in the case of either a perfect or
an imperfect quantum channel. Furthermore, we will show that some recent claims
that the protocol is insecure and that the security proof is wrong, are
erroneous.

\section{The protocol in short}

Bob, the receiver of information, prepares two qubits in the Bell state
$|\Psi^+\rangle_{ht}=\frac1{\sqrt2}(|01\rangle_{ht}+|10\rangle_{ht})$, where
$h$ and $t$ refer to the home and travel qubit, respectively. He sends the
travel qubit to Alice who randomly selects either message mode or control mode.
In message mode, Alice applies the encoding operation $Z_t(j)=\sigma_z^{j}$ to
the travel qubit, where $j\in\{0,1\}$ represents the message bit. For $j=1$,
the encoding operation transforms $|\Psi^+\rangle_{ht}$ into
$|\Psi^-\rangle_{ht}$ and for $j=0$ the state is left unchanged. After
encoding, Alice sends the qubit back to Bob, who then applies a Bell measurement on
both qubits, yielding either $|\Psi^+\rangle_{ht}$ or $|\Psi^-\rangle_{ht}$,
thereby revealing the message bit $j$. In control mode, Alice measures the
travel qubit in the $z$-basis and announces the result via the public channel.
Upon receiving the announcement, Bob measures his home qubit in the $z$-basis
and compares both results. If they differ, the protocol is continued, otherwise
aborted.

\section{Cai's DOS-attack}

Qing-yu Cai~\cite{Cai2003} has proposed a simple attack scheme that disturbs the
information transmission without being detectable and without revealing any message information, that is, a \emph{denial-of-service} (DoS) attack: Eve measures every qubit travelling from Alice to Bob in the $z$-basis. According to the protocol, there is no security check for qubits travelling back from Alice to Bob, hence the attack remains undetected. 
As the attack destroys the entanglement between home and travel qubit, the bit read out by Bob is completely uncorrelated with the bit encoded by Alice, so the message is scrambled. Since Eve's measurement result is completely random, she does not gain any message information. 

Let us point out that a trivial modification of the attack even allows Eve to
derministically change the information transmitted on the channel by flipping
message bits of her choice: Instead of measuring the qubit, Eve applies a
$\sigma_z$-operation. This way, Eve is able to alter message information,
albeit in a ``blind'' way. Altogether, the protocol in its original form
protects the \emph{confidentiality} of the message but not the
\emph{integrity} (just like the classical one-time pad).

Cai himself has proposed fairly simple ways to protect the protocol against
this type of attack: either quantum mechanically, by slightly modifying the
control mechanism, or classically, by performing one of the standard methods of
message authentification~\cite{Schneier1996}.

\section{Wojcik's attack on a lossy quantum channel}

So far, the ping-pong protocol is provably secure only for the case of a perfect quantum channel.
Any imperfection of the channel potentially opens the door to effective and undetectable eavesdropping.
Since all quantum cryptographic protocols are confronted with such a problem, the standard procedure is to introduce additional steps of error correction and privacy amplification using the public channel to distill an asymptotically perfectly secure key. These procedures can be successful exactly if the mutual information between sender and receiver is greater than that between sender and eavesdropper~\cite{CsiszarKorner1978}.

Antoni Wojcik~\cite{Wojcik2003} proposed a smart eavesdropping attack scheme that works on a lossy quantum channel and enables the eavesdropper to gain message information without being detected. 

The basic idea is the following. After receiving the travel qubit from Bob, Eve appends an ancilla system ${\cal H}_{xy}$ in some initial state $|e_0\rangle_{xy}$ to the system ${\cal H}_{ht}$ of travel qubit and home qubit in its initial state, yielding the state 
\begin{equation}
	|\text{init}\rangle=\frac1{\sqrt2}(|01\rangle+|10\rangle)_{ht}|e_0\rangle_{xy}. \label{W1}
\end{equation}
She then unitarily transforms this initial state into
\begin{equation}
	\begin{split}
	|\text{B--A}\rangle=&\frac1{2}|0\rangle_h\big(|1\rangle_t|e_1\rangle_{xy}
		+|{\rm vac}\rangle_t|e_2\rangle_{xy}\big)\\
		&+\frac1{2}|1\rangle_h\big(|0\rangle_t|e_3\rangle_{xy}
		+|{\rm vac}\rangle_t|e_4\rangle_{xy}\big),
	\end{split}
\end{equation}
where $|e_1\rangle$,\ldots,$|e_4\rangle$ are mutually orthogonal. Then Eve resends the travel qubit to Alice, whose encoding operation $\sigma_z^j$ on the travel qubit yields
\begin{equation}
	\begin{split}
	|\text{B--A}'\rangle=&\frac1{2}|0\rangle_h\big((-1)^j|1\rangle_t|e_1\rangle_{xy}
		+|{\rm vac}\rangle_t|e_2\rangle_{xy}\big)\\
		&+\frac1{2}|1\rangle_h\big(|0\rangle_t|e_3\rangle_{xy}
		+|{\rm vac}\rangle_t|e_4\rangle_{xy}\big).
	\end{split}
\end{equation}
It can directly be seen from the above terms that upon capturing the travel qubit being sent from Alice to Bob, Eve is able to find the message bit $j$ with probability $1/2$ by measuring the $txy$ system in a suitable basis or, equivalently, by performing some unitary operation on the $txy$ system and then measuring in the computational basis. In a control run the travel mode is found to be in the vacuum state with probability 1/2. Hence, Eve's attack introduces 50\% channel losses if she attacks all the time. 

If the efficiency of the channel is $\eta<0.5$ then Eve replaces the lossy
channel with a better one, so that the channel exactly mimics the losses expected by Alice and Bob. That way, she can attack all transmissions while staying undetectable, and she creates mutual information between herself and Alice that exceeds the mutual information between Alice and Bob. Hence, even with error correction and privacy
amplification, the protocol would not be secure. If $0.5<\eta\leq0.6$, she
attacks the fraction $\mu=2(1-\eta)$ of qubits. For efficiencies above 0.6, the
mutual information between Alice and Eve falls below the mutual information
between Alice and Bob, so that error correction and privacy amplification can
establish the security of the protocol.

Fortunately, Wojcik himself proposed in the same paper two solutions to protect the protocol
against his attack. One solution is to estimate the qubit error rate (QBER)
which, however, forces Alice and Bob to sacrifice some of their message bits.
The other solution is to delay Alice's announcement of the transmission mode
(message or control), until Bob has checked if there is an additional photon in
the travel mode. This way, the attack can be detected because the particular attack
operation not only produces channel losses, but also (with probability 1/2)
inserts a photon into the travel mode which in control mode should be in vacuum
state after Alice's measurement. Such an ``illegal'' photon travelling to Bob
would, if detected, immediately reveal the presence of the eavesdropper.

In summary, Wojcik's attack exploits a security hole of the protocol in the
realistic case of an imperfect quantum channel using photons. A fairly
simple modification of the protocol closes the hole and restores the security
of the protocol.

\section{The ZML-attack on an imperfect quantum channel}

In~\cite{ZhangMan2004}, Zhan-jun Zhang, Zhong-xiao Man, and Yong Li improved W\'ojcik's attack by expanding the domain of effective eavesdropping from nearly 60\% to nearly 80\% channel efficiency. 
The basic idea is fully analoguous to W\'ojcik's scheme~\cite{Wojcik2003}. Eve appends an ancilla system ${\cal H}_{xy}$ and unitarily transforms the state~\eqref{W1} into the state
\begin{equation}
	\begin{split}
	|\text{B--A}\rangle=&\frac1{2}|0\rangle_h\big(|1\rangle_t|e_1\rangle_{xy}
		+|{\rm vac}\rangle_t|e_2\rangle_{xy}\big)\\
		&+\frac1{\sqrt2}|1\rangle_h\big(|0\rangle_t|e_3\rangle_{xy}\big),
	\end{split}
\end{equation}
where $|e_1\rangle$,\ldots,$|e_4\rangle$ are mutually orthogonal. Alice's encoding operation on the travel qubit yields
\begin{equation}
	\begin{split}
	|\text{B--A}'\rangle=&\frac1{2}|0\rangle_h\big((-1)^j|1\rangle_t|e_1\rangle_{xy}
		+|{\rm vac}\rangle_t|e_2\rangle_{xy}\big)\\
		&+\frac1{\sqrt2}|1\rangle_h\big(|0\rangle_t|e_3\rangle_{xy}\big).
	\end{split}
\end{equation}
Again, by measuring the $txy$ system Eve finds the message bit $j$ with probability $1/2$. In a control run, the travel mode is found to be in the vaccum state with probability 1/4. Hence, Eve's attack introduces 25\% channel losses if she attacks all the time. Reasoning analoguous to that in W\'ojcik's publication reveals that the ping-pong protocol can be successfully attacked for channel efficiencies up to nearly 80\%. However, also here the security of the protocol can be re-established by introducing the same countermeasures suggested by W\'ojcik.

\section{The ZLM-attack on a perfect quantum channel}

Recently, the same three authors proposed an attack scheme against the ping-pong protocol, enabling the eavesdropper to read out message information without being detected even in the
case of a perfect quantum channel, in contradiction to our rigorous security
proof for this case~\cite{ZhangLi2005}. However, their attack scheme is faulty:

According to their attack scheme, Eve prepares an ancilla state $|\chi\rangle=|{\rm vac},0\rangle_{xy}$ in two additional modes $x$ and $y$, and applies a unitary operation $W_{txy}$ (Eq.~(2) in \cite{ZhangLi2005}) on the compound system $txy$ of the travel qubit and the ancilla modes during the B-A-transmission.
Afterwards, the total system is in the state
\begin{equation}
	\begin{split}
	|B-A\rangle
		&=\frac12|0,1\rangle_{ht}(|{\rm vac}, 0\rangle_{xy}+|1,{\rm vac}\rangle_{xy})\\
		&\quad+\frac12|1,0\rangle_{ht}(|{\rm vac}, 1\rangle_{xy}+|0,{\rm vac}\rangle_{xy}).
	\end{split}
\end{equation}
It is clear that this attack operation cannot be detected by the control
measurements of the ping-pong protocol: $z$-basis measurements on $h$ and $t$
will still be strictly anticorrelated. 

In message mode, Alice applies the encoding operation $Z_t^{j}=\sigma_z^{j}$ to
the travel photon, where $j\in\{0,1\}$ represents the message bit, and sends
the photon back to Bob.
Eve intercepts the travel photon, applies the inverse operation $W_{txy}^{-1}$ on the compound system $txy$, resends the travel photon to Alice and keeps her ancilla system.
The authors claim that a measurement on the ancilla system reveals information about the message bit $j$ encoded by Alice. Indeed, the authors' Eq.~(7), which supposedly shows the state $|A-B\rangle$ after Eve's (A-B)-attack operation $W_{txy}^{-1}$, indicates that the message bit $j$ is partly encoded in the state of the $y$ photon: 
\begin{equation}
	\begin{split}\label{Eq7}
		|A-B\rangle_j 
			&= \frac12\Big[(-1)^j(\Psi_{ht}^++\Psi_{ht}^-)|j\rangle_y\\
			&\quad+(\Psi_{ht}^+-\Psi_{ht}^-)|0\rangle_y\Big]|{\rm vac}\rangle_x.
	\end{split}
\end{equation}

Obvously, a computational-basis measurement by Eve on the $y$-mode reveals the message bit $j$ with probability 1/2, otherwise it yields 0. However, this crucial equation is wrong, which can be seen as follows. When Alice applies her encoding operation to the travel photon $t$, the total system is in the state
\begin{equation}
	\begin{split}
	Z_t^{j}|B-A\rangle
		&=\frac1{\sqrt2}\Big[(-1)^j|0,1\rangle_{ht}|\chi_1\rangle_{xy}
		+|1,0\rangle_{ht}|\chi_0\rangle_{xy}\Big],
	\end{split}
\end{equation}
where we have set
\begin{eqnarray}
	|\chi_1\rangle_{xy}&=&\frac1{\sqrt2}(|{\rm vac},0\rangle_{xy}+|1,{\rm vac}\rangle_{xy})\\
	|\chi_0\rangle_{xy}&=&\frac1{\sqrt2}(|{\rm vac},1\rangle_{xy}+|0,{\rm vac}\rangle_{xy}).
\end{eqnarray}
As can be seen from above, the message bit $j$ is encoded in the relative phase between the two components of the superposition. Since Eve has no access to the home photon $h$, she can in no way read out the relative phase.
Generally, consider a state in the space ${\cal H}_{h}\otimes {\cal H}_E$ of the form
\begin{equation}
	|\Psi\rangle = \alpha|h_1\rangle_h|e_1\rangle_E+\beta e^{i\phi}|h_2\rangle_h|e_2\rangle_E,
\end{equation}
with $\langle h_1|h_2\rangle=0$, and $\alpha,\beta>0$, and where Eve has only access to the system in ${\cal H}_E$. Then for Eve the state of the entire system is indistinguishable from the reduced density matrix
\begin{eqnarray}
	\rho_E&=&{\rm Tr}_h\{|\psi\rangle\langle\psi|\}\\
		&=&\alpha^2|e_1\rangle\langle e_1|_E+\beta^2|e_2\rangle\langle e_2|_E,
\end{eqnarray}
where the relative phase $\phi$ is no longer available.

In the present case, the density matrix of the total state after Alice's encoding operation is
\begin{eqnarray}
	\begin{aligned}
	\rho_j & = Z_t^j|B-A\rangle\langle B-A|Z_t^{j\dagger}
\cr	& = \frac12\Big[|01\rangle_{ht}|\chi_1\rangle_{xy}\langle01|_{ht}\langle\chi_1|_{xy}\\
		&\quad+(-1)^j|01\rangle_{ht}|\chi_1\rangle_{xy}\langle10|_{ht}\langle\chi_0|_{xy}\\
		&\quad+(-1)^j|10\rangle_{ht}|\chi_0\rangle_{xy}\langle01|_{ht}\langle\chi_1|_{xy}\\
		&\quad+|10\rangle_{ht}|\chi_0\rangle_{xy}\langle10|_{ht}\langle\chi_0|_{xy}\Big].
	\end{aligned}
\end{eqnarray}
The state of the system accessible to Eve is given by partial-tracing over the home photon $h$,
\begin{equation}
	\rho_{j}^{({\rm Eve})}={\rm Tr}_h\{\rho_j\},
\end{equation}
which yields
\begin{equation}
	\begin{split}
	\rho_{j}^{({\rm Eve})} 
		&= \frac12\Big[|1\rangle_{t}|\chi_1\rangle_{xy}\langle1|_{t}\langle\chi_1|_{xy}\\
		&\quad+|0\rangle_{t}|\chi_0\rangle_{xy}\langle0|_{t}\langle\chi_0|_{xy}\Big].
	\end{split}
\end{equation}

Since $\rho_j^{\rm (Eve)}$ is independent of $j$, there is no message information available to Eve.
Consequently, the authors' calculation of the state $|A-B\rangle_j$, which results from the application of $W_{txy}^{-1}$ to $Z_t^j|B-A\rangle$ must be faulty. In fact, our own calculations show that the state after Eve's (A-B)-attack $W_{txy}^{-1}$ reads
\begin{equation}
	\begin{split}
		|A-B\rangle_j 
			&= \frac12\Big[(-1)^j(\Psi_{ht}^++\Psi_{ht}^-)\\
			&\quad+(\Psi_{ht}^+-\Psi_{ht}^-)\Big]|0,{\rm vac}\rangle_{xy},
	\end{split}
\end{equation}
which is different from the authors' Eq.~(7)~(Eq.~\eqref{Eq7} above) in that it is impossible for Eve to read out the message bit by any measurement performed on the ancilla system $xy$.

In summary, we find that the conclusions drawn in the commented paper \cite{ZhangLi2005} are based on a miscalculation; this attack scheme is not effective and does not impair the security of the ping-pong protocol.

\section{Cai's invisible photon attack}

Qing-Yu Cai~\cite{Cai2006} adapted the Trojan horse attack introduced by Gisin
et al~\cite{GisinFasel2005} to the ping-pong protocol: Eve feeds in an
additional photon which is invisible to Alice and Bob's detectors, but which is
affected by Alice's encoding operation. The illegal photon is inserted into the
travel mode on the way from Bob to Alice, and it is filtered out during the
transmission from Alice to Bob. Eve detects the state change of the
illegal photon which is caused by Alice's encoding operation, and thereby
obtains the message bit without being detected. Choosing a wavelength
outside the range of Alice's detectors is one possible way to make the
illegal photon invisible to the control measurements.
As Cai has pointed out, the attack does not exploit a weakness of the protocol itself but rather of certain imperfect implementations of the protocol. He also suggests a feasible solution to re-establish the security of the communication: Alice and Bob add filters to their setup whose bandwidth matches the sensitivity range of the detectors. 

The generalization is straightforward: The experimental setup should block any quantum carriers of information which are invisible to the detectors but which are affected by the encoding operation.

\section{Zhang's claim that the security proof is wrong}

Zhan-jun Zhang challenges the validity of our security proof altogether~\cite{Zhang2006}. We will show that the claim is based
1) on a misunderstanding of the security proof and
2) on a miscalculation at a crucial point in the argument.

The author emphasizes that in our security proof the eavesdropping information $I_0$ is extracted from the travel qubit only.
This is not the case. The maximal amount $I_0$ of information that can be extracted from the system available to Eve, is equal to the von-Neumann information $S$ of the state $\rho''$ given by Eq. (8) in our original paper.
The state $\rho''$ results from Alice's encoding operation on the state $\rho'$ which is given by our Eq.~(7) as a matrix representation in the orthogonal basis $\{|0,\chi_0\rangle, |1,\chi_1\rangle\}$. As we have pointed out in our paper, the states $|\chi_0\rangle$ and $|\chi_1\rangle$ are states of Eve's ancilla system ${\cal H}_E$. It is therefore not true that the information $I_0$ is derived only from the state of the travel qubit. Unfortunately, though, we ourselves have made such misunderstanding easy because right before our Eq.~(8) we denote $\rho''$ as ``the state of the travel qubit after Eve's attack operation and after Alice's encoding operation''. This is a misnomer which we apologize for; it should however be clear from the context that $\rho''$ refers to the state $\rho'$ after Alice's encoding operation, and that the state $\rho'$ explicitely includes Eve's ancilla system.

Based on this misunderstanding, the author constructs a counterexample against the security proof, where he then miscalculates the information contents $I_{0t}$ and $I_{0c}$ that can be extracted from the travel qubit and the composite system, respectively. He claims that ``as can easily be worked out'', the values read $I_{0t}=1$ and $I_{0c}=2$, which would be in contradiction to the prepositions of our security proof.

Let us explicitely perform the calculation. 
According to the author's counterexample, Eve captures the travel qubit $t$ in the state $|0\rangle$, attaches an ancilla system $x$ in the state $|\chi\rangle_x = {1\over\sqrt{2}}(|0\rangle_x+|1\rangle_x)$. We find that the state $|\Psi'\rangle$ of the composite system $tx$ after Eve's
attack operation $\hat E$ given by the author reads
\begin{eqnarray}
	|\Psi'\rangle &=&\hat E (|0\rangle_t |\chi\rangle_x) \\
		&=& \hat E\frac1{\sqrt2}(|00\rangle_{tx}+|01\rangle_{tx})\\
		&=&\frac1{\sqrt2}|0\rangle_t|\chi_0\rangle_x
			+\frac1{\sqrt2}|1\rangle_t|\chi_1\rangle_x,
\end{eqnarray}
where we have set
\begin{eqnarray}
	|\chi_0\rangle_x&=&\frac1{\sqrt2}(|0\rangle_x+|1\rangle_x)\\
	|\chi_1\rangle_x&=&\frac1{\sqrt2}(|1\rangle_x-|0\rangle_x).
\end{eqnarray}

The state $|\Psi'\rangle$ has exactly the form given in Eq.~(4) of our security proof, with $\alpha=\beta=\frac1{\sqrt2}$. When Alice encodes ``0'' she applies the unity operation which gives $|\Psi'_0\rangle=|\Psi'\rangle$; when she encodes ``1'' she applies $\sigma_z$ to the travel qubit which gives
\begin{equation}
	|\Psi'_1\rangle=\frac1{\sqrt2}|0\rangle_t|\chi_0\rangle_x
			-\frac1{\sqrt2}|1\rangle_t|\chi_1\rangle_x.
\end{equation}

Note that $|\Psi'_1\rangle$ is orthogonal to $|\Psi'_0\rangle$. Assuming that she encodes ``0'' or ``1'' with equal probability (which is tacitly assumed by the author), the state of the composite system $tx$ reads
\begin{equation}
	\rho''=\frac12|\Psi'_0\rangle\langle\Psi'_0|
		+\frac12|\Psi'_1\rangle\langle\Psi'_1|,
\end{equation}
which has the entropy $I_{0c}=S(\rho'')=1$, in contrast to the authors' result.
In fact, the author's value of $I_{0c}=2$ would be very surprising: Alice
encoded at most one classical bit by a unitary operation, so the entropy of the
resulting state cannot be higher than one bit. For security reasons, our protocol makes use only of a 2-dimensional subspace of the full 4-dimensional Hilbert space spanned by two qubits, and the entropy of a state in a 2-dimensional subspace is at most 1 bit.

\section{Conclusion}

So far, the ping-pong protocol has resisted all serious attacks brought forward in the last five years, albeit with slight modifications of the scheme. For the ideal case of a perfect quantum channel, the initial security proof holds and is both rigorous and general. 
For the realistic case of an imperfect quantum channel, there is no general security proof but the protocol seems to retain its security.

We suggest that future efforts should go into either figuring out more attack scenarios exploiting channel imperfections under realistic circumstances or into finding a general proof for the unconditional security of (a suitable extension of) the ping-pong protocol in the case of an imperfect quantum channel.

\end{document}